\documentclass[letterpaper]{article} % DO NOT CHANGE THIS
\usepackage{aaai24}  % DO NOT CHANGE THIS
\usepackage{times}  % DO NOT CHANGE THIS
\usepackage{helvet}  % DO NOT CHANGE THIS
\usepackage{courier}  % DO NOT CHANGE THIS
\usepackage[hyphens]{url}  % DO NOT CHANGE THIS
\usepackage{graphicx} % DO NOT CHANGE THIS
\usepackage{natbib}  % DO NOT CHANGE THIS AND DO NOT ADD ANY OPTIONS TO IT
\usepackage{caption} % DO NOT CHANGE THIS AND DO NOT ADD ANY OPTIONS TO IT
\usepackage{algorithm}
\usepackage{algorithmic}
\usepackage{amsmath}
\usepackage{amsthm}
\theoremstyle{definition}
\newtheorem{definition}{Definition}
\newtheorem{proposition}{Proposition}
\usepackage{subcaption}
\usepackage{amssymb}
\usepackage{tabularx}
\usepackage{booktabs}
\usepackage{multirow}
\usepackage{amsmath}
\usepackage{amsthm}

\usepackage{pdfpages} % include pdfs
\usepackage{pgffor} % for loops

\makeatletter
\AtBeginDocument{\let\LS@rot\@undefined}
\makeatother

\def\supplementfilename{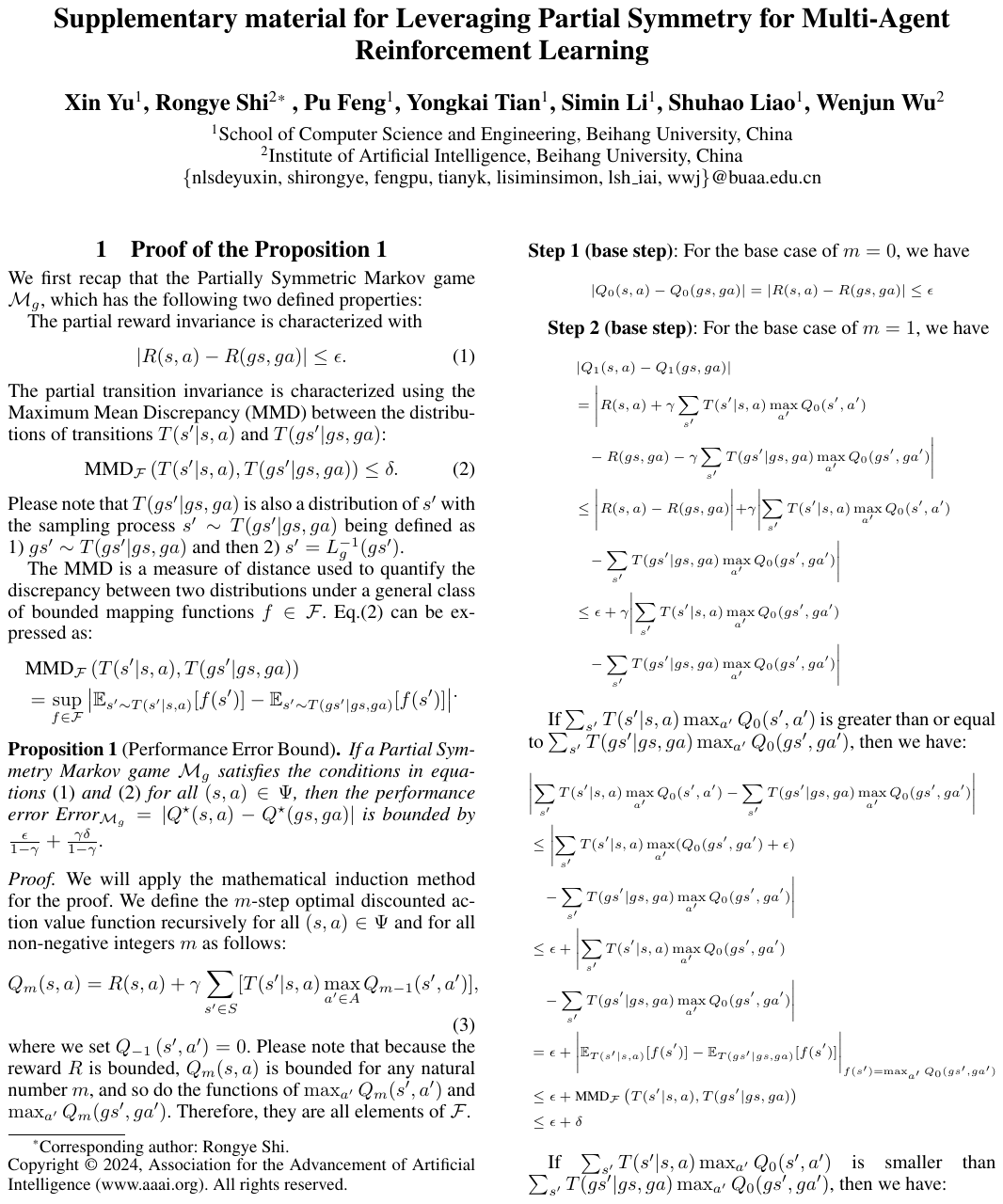}

\pdfximage{\supplementfilename}
\def\numbersupplementpages{\the\pdflastximagepages}

\newif\ifarXiv
\arXivtrue 
\usepackage{newfloat}
\usepackage{listings}
\DeclareCaptionStyle{ruled}{labelfont=normalfont,labelsep=colon,strut=off} % DO NOT CHANGE THIS
\floatstyle{ruled}
\newfloat{listing}{tb}{lst}{}
\floatname{listing}{Listing}
\title{Leveraging Partial Symmetry for Multi-Agent Reinforcement Learning}
\author{
    Xin Yu\textsuperscript{\rm 1},
    Rongye Shi\textsuperscript{\rm 2}\thanks{Corresponding author: Rongye Shi.} ,
    Pu Feng\textsuperscript{\rm 1},
    Yongkai Tian\textsuperscript{\rm 1},
    Simin Li\textsuperscript{\rm 1},
    Shuhao Liao\textsuperscript{\rm 1},
    Wenjun Wu\textsuperscript{\rm 2}
}

\affiliations{
    \textsuperscript{\rm 1}School of Computer Science and Engineering, Beihang University, Beijing, China\\
    \textsuperscript{\rm 2}Institute of Artificial Intelligence, Beihang University, Beijing, China\\
     \{nlsdeyuxin, shirongye, fengpu, tianyk, lisiminsimon, lsh\_iai, wwj\}@buaa.edu.cn    
}

\begin{document}

\maketitle

\begin{abstract}
Incorporating symmetry as an inductive bias into multi-agent reinforcement learning (MARL) has led to improvements in generalization, data efficiency, and physical consistency. While prior research has succeeded in using perfect symmetry prior, the realm of partial symmetry in the multi-agent domain remains unexplored. To fill in this gap, we introduce the partially symmetric Markov game, a new subclass of the Markov game. We then theoretically show that the performance error introduced by utilizing symmetry in MARL is bounded, implying that the symmetry prior can still be useful in MARL even in partial symmetry situations. Motivated by this insight, we propose the Partial Symmetry Exploitation (PSE) framework that is able to adaptively incorporate symmetry prior in MARL under different symmetry-breaking conditions. Specifically, by adaptively adjusting the exploitation of symmetry, our framework is able to achieve superior sample efficiency and overall performance of MARL algorithms. Extensive experiments are conducted to demonstrate the superior performance of the proposed framework over baselines. Finally, we implement the proposed framework in real-world multi-robot testbed to show its superiority.
\end{abstract}

\section{Introduction}
Multi-Agent Reinforcement Learning (MARL) is increasingly gaining attention due to its capabilities in handling complex tasks\cite{li2023byzantine}. Such tasks often necessitate strategic interaction and rivalry among various entities~\cite{yu2021swarm,feng2023mact}. However, a notorious limitation of most MARL approaches is their substantial data dependency, necessitating vast amounts of data to build an efficient model. This limitation severely narrows the scope of MARL's practical application in real-life settings~\cite{shi2021improving}. How to develop strategies to improve MARL's sample efficiency has become an important and long-standing research topic.

\begin{figure}[ht]
\centerline{\includegraphics[width=5.5cm]{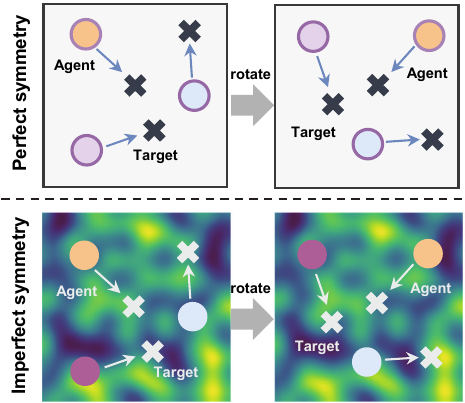}}
\caption{Illustration of symmetry disruption in a non-uniform field: despite the spatial symmetry of the multi-agent system, the introduction of a non-uniform field, such as uneven terrain or a wind field, disrupts this symmetry and the symmetry assumption does not strictly hold everywhere. Colors denote varying intensities of the field.}
\label{example}
\end{figure}

Strategies to augment sample efficiency often involve integrating external knowledge to accelerate MARL's training. Various methods incorporating extra knowledge have been proposed in recent literature~\cite{shi2021physics,shi2021tits}. In the realm of MARL, prior research has highlighted the advantage of employing permutation invariance. Permutation invariance asserts that the systemic behavior remains unaffected by any changes in the order of agent consideration~\cite{api}. The application of permutation invariance encourages extensive parameter sharing among agents, thereby augmenting data efficiency. Additionally, the most common symmetry in multi-agent systems is rotation symmetry, as illustrated in Figure~\ref{example}. In this context, rotating the global state results in a rotation of the optimal joint policy. Some studies have ameliorated data efficiency by designing inherent network structures that satisfy this property~\cite{vand}.

Current techniques often assume the existence of perfect permutation invariance or perfect spatial symmetry. However, such ideal conditions are rare in real-world scenarios. For instance, again in Figure~\ref{example}, multiple agents attempt to approach a target point where each agent can sense the environment, including information about other agents, obstacles, and the target point. Such problems, conditioned on the perfect symmetry transition function and symmetry reward function, are defined as symmetric Markov game in \cite{vand,yu2023esp}. Unfortunately, in the real world, there might exist imperfections in the environment, e.g., uneven ground, wind, and other non-uniform fields acting on the agents. The non-uniform fields can deviate the system's transition dynamics or reward functions from perfect spatial symmetry to a certain extent. Specifically, when we rotate the state-action pairs of our agents, we cannot rotate the non-uniform fields, accordingly. As a result, despite the multi-agent system having a spatially symmetrical structure, its response to the non-uniform fields can no longer employ perfect symmetry. Furthermore, slight variations in elements such as power supply or physical structures could make the agents slightly heterogeneous, thus violating the principle of perfect permutation invariance. This violation poses a significant challenge to real-world implementations of symmetry-prior-based reinforcement learning methods. 

\begin{figure}[ht]
\centerline{\includegraphics[width=6.5cm]{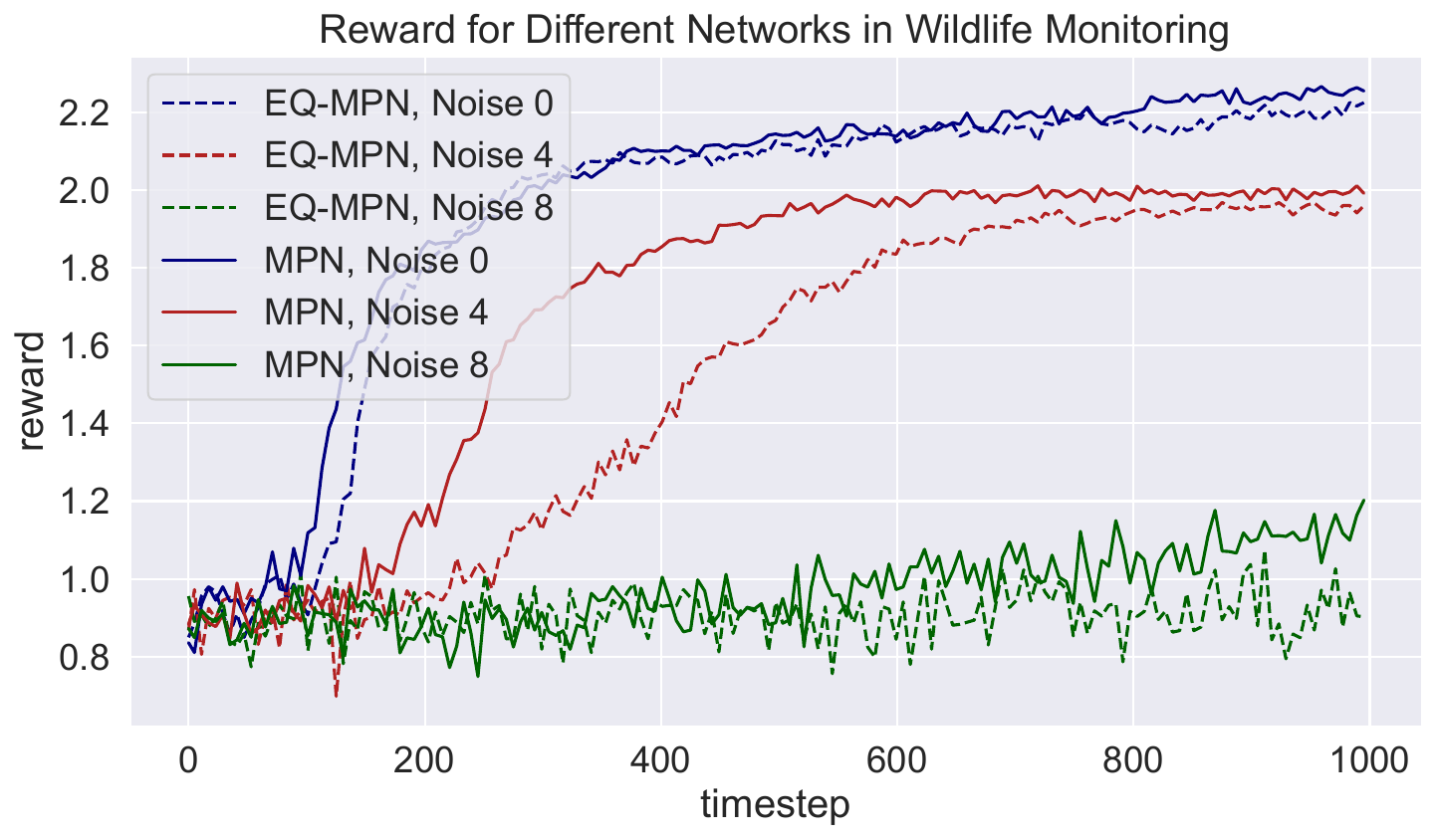}}
\caption{Performance of the EQ-MPN and MPN under varying noise levels. As the noise intensity (the degree of symmetry-breaking) increases, the perfect symmetry network, EQ-MPN, exhibits a declining trend.}
\label{motivation}
\end{figure}

Regrettably, existing studies, either single-agent or multi-agent ones, have seldomly explored such scenarios of partial symmetry, neither from a theoretical nor from a practical point of view. To emphasize the necessity of such a study, we evaluated the performance of the perfect symmetry network proposed in \cite{vand} under various symmetry-breaking conditions. As depicted in Figure~\ref{motivation}, the performance of their EQ-MPN and MPN methods is evaluated under three distinct noise levels, which signifies the extent of symmetry-breaking introduced into the system. We found that as symmetry breaks, the performance of the network with embedded symmetry, EQ-MPN, deteriorates. Motivated by these challenges, we delve into the partial symmetry scenarios, targeting at a new methodology that relaxes the requirement of strict symmetry with a theoretical performance bound guaranteed.

In this paper, we first define the partially symmetric Markov game. It gives rise to a new class of symmetry Markov game with slack symmetry constraints while partially maintaining favorable inductive biases for learning. We theoretically show that the performance errors introduced by leveraging symmetry under partially symmetric Markov game are bounded. Our theoretical analysis can be seamlessly applied to a variety of symmetries, including permutation invariance, rotational equivariance, etc. Upon this setting, we introduce a Partial Symmetry Exploitation framework (PSE). PSE first quantifies the extent/level of symmetry in the environment using a dedicated symmetry quantification module and then selects an appropriate training pipeline according to that symmetry level. The PSE involves several technical components to adaptively incorporate symmetry into the training process. Our main contributions are listed as follows:
\begin{itemize}
\item Formally define the concept of partial equivariance and generalize symmetry Markov game to partially symmetric Markov game;
\item For partially symmetric Markov game, theoretically show that the performance error introduced by utilizing symmetry in MARL is bounded;
\item Motivated by the error bound, propose a novel PSE framework to adaptively incorporate and leverage symmetry prior in MARL;
\item Demonstrate our framework's superiority over baselines in both simulated tasks and real-world robot experiments.
\end{itemize}

\section{Related work}
\subsection{Symmetries in Single-agent RL}

The methods for exploiting symmetry in RL can be broadly classified into two major categories: data augmentation and network structure design. Data augmentation in single-agent RL is to generate additional data through image transformations during the training phase of the model~\cite{rlaug,imageallneed,ITER,arm}. Alternatively, symmetry can be introduced through a contrastive learning framework by enforcing consistencies between an image and its augmented version ~\cite{curl}. The network design method is to design specialized architectures that implicitly embed prior knowledge relevant to the task~\cite{report}. For instance, symmetries in the joint state-action space can be expressed through the implementation of policy networks~\cite{homon,so2}. Our paper explores the realm of partial symmetry, extending beyond the scope of the approaches commonly employed.

\subsection{Symmetries in Multi-agent RL}

In the realm of multi-agent systems, fewer studies have explored the use of data augmentation techniques. To our knowledge, the most related work to our work is the data augmentation method proposed in~\cite{homoaug}. This method generates additional data by implementing permutation transformations for homogeneous agents, interpreting data augmentation from the perspective of permutation invariance. In a similar vein, the need for more extensive integration of prior knowledge into MARL is apparent. Multi-Agent MDP Homomorphic Networks have been developed to embed symmetries, thus enhancing data efficiency~\cite{vand}. However, these methods impose strict constraints on symmetry, which hinders their applicability in real-world scenarios characterized by partial symmetry. In contrast, we treat symmetry as an additional objective and incorporate it through soft constraints such as data augmentation and regularization. Our approach is able to adjust to different symmetry levels, thereby improving algorithmic performance in scenarios with partial symmetry.

\section{Preliminaries}

\subsection{Cooperative Markov game}
An $n$-agent cooperative Markov game \cite{Boutilier96} can be defined as a tuple $(N,S,\left\{A_{i}\right\}_{i=1}^{n}, R, T,\Psi)$, where $N$ denotes the set of agents, $S$ is the state space, and $A_{i}$ is the action space of agent $ i=1, \ldots, n $. Let $A=A_{1} \times A_{2} \times \cdots \times A_{n}$ be the joint action space, and $T: S \times A \times S \rightarrow[0,1]$ be the transition function. $\Psi$ is the set of admissible state-action pairs. 
At time step $t$, the agents are at state $s_t$ (which may not be fully observable) and take independent action $(a_1,..., a_N)$ relying on their policy. Then, the environment emits the bounded joint reward $R$ and moves to the next state $s_{t+1}$.
The agents aim to maximize the expected joint return, defined as $\mathbb{E}_{\pi}\left[\sum_{t=0}^{\infty} \gamma^t R\left(s_t, a_t\right)\right]$, where $0<\gamma<1$ is the discount factor, by selecting actions according to the policy $\pi_{i}:{S} \times{A}_{i} \rightarrow[0,1]$. The initial states are determined by a distribution $\eta:{S} \rightarrow[0,1]$.

\subsection{Groups and Transformations}
This section offers an overview of the concepts of groups and transformations~\cite{bros}. A group $G$ is a set equipped with a binary operator that has four mathematical properties: identity, inverse, closure, and associativity. Our discussion primarily revolves around the group $ \mathrm{SO}(2) $ and its cyclic subgroup $ C_n $. Specifically, $ \mathrm{SO}(2) $ represents the group of continuous rotations  $ \left\{\mathrm{R}_\theta: 0 \leq \theta < 2 \pi \right\} $. Meanwhile, $ C_n $ stands for the discrete subgroup, defined as $ C_n=\left\{\mathrm{R}_\theta: \theta \in \left\{\frac{2 \pi i}{n} \mid 0 \leq i < n \right\} \right\} $. A rotation matrix illustrates the act of rotating within Euclidean space~\cite{rotationm}.  For a specific rotation set $\left\{0^{\circ}, 90^{\circ}, 180^{\circ}, 270^{\circ}\right\}$, the rotation matrix is formulated as:
\begin{equation*}
    R(\theta)=\left[\begin{array}{cc}\cos \theta & -\sin \theta \\ \sin \theta & \cos \theta\end{array}\right]\label{rotation1}.
\end{equation*}
The four group axioms are satisfied in the case of a rotation transformation.

\subsection{Equivariance and Invariance}

In multi-agent systems, the symmetries are commonly referred to as equivariance and invariance\cite{yu2023esp}. Given a transformation operator $L_g: \mathcal{X} \rightarrow \mathcal{X}$ and a mapping function $f: \mathcal{X} \rightarrow \mathcal{Y}$, if there exists a second transformation operator $K_g: \mathcal{Y} \rightarrow \mathcal{Y}$ in the output space of $f$ such that:
\begin{equation*}
    K_g[f(x)]=f\left(L_g[x]\right),
\end{equation*}
where $g \in G$ and $G$ is a mathematical group, then, function $f$ is equivariant to the transformation $L_g$. The operators $L_g$ and $K_g$ can be used to describe the same transformation, but in different spaces. A related notion to equivariance is invariance. If for any choice of $g \in G$, we have that $K_g=I$, the identity function, then we say function $f$ is invariant to transformation $L_g$. Figure \ref{example} (upper half) shows the equivariance of the optimal policy, rotating the state globally results in a transformation of the optimal policy. Given two states $s$ and $L_g[s]$, the optimal policy $\pi^*$ is equivariant to its transformation which is denoted by $ K_g[\pi^*(s)]=\pi^*\left(L_g[s]\right)$. Without special notice, the transformations $L_g, K_g$ are assumed to be \textit{bijective} in this paper.

\section{Defining and Characterizing Partially Symmetric Markov game}
\subsection{Partial Equivariance and Invariance}

Real-world dynamics may not satisfy the strict equivariance though such a strict assumption has been commonly used for simplicity in literature~\cite{so2,homon}. In this paper, we introduce a definition of partial equivariance and invariance to fill the gap.
\begin{definition}[Partial Equivariance and Invariance]
Given a transformation operator $L_g: \mathcal{X} \rightarrow \mathcal{X}$ and a mapping function $f: \mathcal{X} \rightarrow \mathcal{Y}$, if there exists a second transformation operator $K_g: \mathcal{Y} \rightarrow \mathcal{Y}$ in the output space of $f$ such that:
\begin{equation*}
   \left \| K_g[f(x)]-f\left(L_g[x]\right) \right \| \leq \epsilon \label{equimp},
\end{equation*} 
where $g \in G$ and $G$ is a mathematical group, we say $f$ is $\epsilon$-partially equivariant to the transformation $L_g$ and $K_g$. A related notion to $\epsilon$-partial equivariance is the $\epsilon$-partial invariance: If for any choice of $g \in G$ we have  $K_g=I$, then function $f$ is $\epsilon$-partially invariant to transformation $L_g$. Note that strict equivariance or invariance are special cases of partial ones with $\epsilon = 0$.
\end{definition}

\subsection{Partially Symmetric Markov game}
In this subsection, we formally define the partially symmetric Markov game, a subclass of the cooperative Markov game characterized by partial symmetry.

\begin{definition}[Partially Symmetric Markov game]
The partially symmetric Markov game $\mathcal{M}_g=(N, S,  \left\{A_{i}\right\}_{i=1}^{n}, R, T, \Psi, g, \epsilon,  \delta )$ is a cooperative Markov game that satisfies the conditions of partial reward invariance and partial transition invariance. The state and action transformation are defined as $L_{g}: S \rightarrow S$ and $K_{g}: A \rightarrow A$, respectively. For state-action pairs $(s,a) \in \Psi$, we denote the transformed state-action pairs as $(gs,ga)$. Here, $
gs=L_{g}(s)$ and $ga=K_{g}(a)$ for short. The partial reward invariance is characterized with:
\begin{equation}
    |R(s, a) - R(gs, ga)| \leq \epsilon \label{rewardequ}.
\end{equation}
The partial transition invariance is characterized using the Maximum Mean Discrepancy (MMD) between the distributions $T(s'|s, a)$ and $T(gs'|gs, ga)$: 
\begin{equation}
    \text{MMD}_{\mathcal{F}}\left(T(s'|s, a), T(gs'|gs, ga)\right)\leq \delta \label{transequ}.
\end{equation}
Please note that $T(gs'|gs, ga)$ is also a distribution of $s'$ with the sampling process $s' \sim T(gs'|gs, ga)$ being defined as 1)~$gs' \sim T(gs'|gs, ga)$ and then 2) $s'=L_g^{-1}(gs')$.

The MMD is defined as a measure of distance used to quantify the discrepancy between two distributions under a general class of bounded mapping functions $f \in \mathcal{F}$~\cite{MMD}. Eq.(2) can be expressed as:
\begin{equation*}
\begin{aligned}
    &\text{MMD}_{\mathcal{F}}\left(T(s'|s, a), T(gs'|gs, ga)\right) \\
    &= \sup_{f \in \mathcal{F}} \left| \mathbb{E}_{s' \sim T(s'|s, a)}[f(s')] - \mathbb{E}_{s' \sim T(gs'|gs, ga)}[f(s')] \right| \\   
\end{aligned}.
\end{equation*}
\end{definition}
We believe that the symmetry property is a common occurrence in many real-world multi-agent tasks. Our formal definition of this issue holds theoretical and practical significance.

\subsection{Performance Error Analysis}

\begin{figure*}[!t]
\centerline{\includegraphics[width=14.5cm]{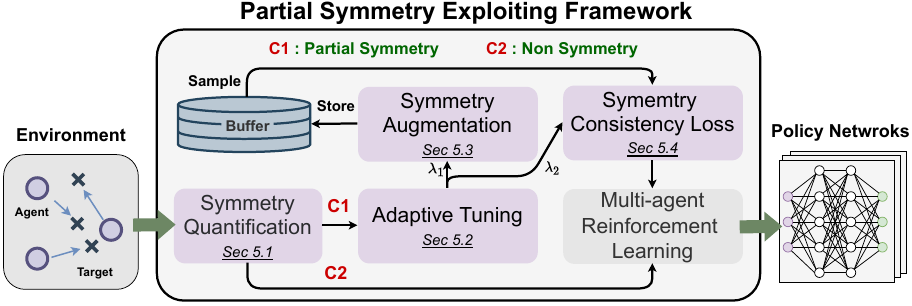}}
\caption{The overall framework of the proposed PSE. The framework is composed of four key modules: 1) Symmetry Quantification, which measures the level of symmetry in the environment, 2) Adaptive Tuning, which serves as the annealing coefficient modulating the continuous degree of symmetry utilization. 3) Symmetry Augmentation, which manipulates the data based on the quantified symmetry, and 4) Symmetry Loss, a specially crafted function that optimizes the policy network with respect to the symmetry.}
\label{framework}
\end{figure*}

We show that if a problem can be formulated as a partially symmetric Markov game, the performance error introduced by using symmetry-augmented data in training is bounded. In the following, variables without a subscript $i$ denote the concatenation of all variables for all agents (e.g., $a$ denotes the joint actions of all agents). Based on the definition in \cite{report}, the $m$-step optimal discounted action value function recursively for all $(s, a) \in \Psi$ and for all non-negative integers $m$ is defined as follows:
% \begin{equation}  
% \begin{aligned} 
%     &Q_{m}(s, a) \\
%     &=R(s, a)+\gamma \sum_{s^{\prime} \in S}[T(s, a, s^{\prime}) \max _{a^{\prime} \in A} Q_{m-1}(s^{\prime}, a^{\prime})]. \label{eq1}
% \end{aligned}
% \end{equation}
\begin{equation*}  
\begin{aligned} 
    &Q_{m}(s, a)=R(s, a)+\gamma \sum_{s^{\prime} \in S}[T(s'|s, a) \max _{a^{\prime} \in A} Q_{m-1}(s^{\prime}, a^{\prime})]. \label{eq1}
\end{aligned}
\end{equation*}
The optimal action-value function $Q^*(s, a)$ is the limit of $Q_m(s, a)$ as $m$ approaches infinity. We now define the performance error for $\mathcal{M}_g$, which measures the error of a Q-function trained on symmetry-augmented data.
\begin{definition}[Performance Error for $\mathcal{M}_g$ when using symmetry-augmented data]
Let $Q^*(s, a)$ be the optimal action-value function, $g$ be the transformation associated with $\mathcal{M}_g$. Then, the performance error of $\mathcal{M}_g$ is defined as:
\begin{equation*}
\textit{Error}_{\mathcal{M}_g}=|Q^{\star}(s, a) - Q^{\star}(gs, ga)|,
\end{equation*}
where $Q^{\star}(gs, ga)$ is the action-value function trained by the symmetry-augmented data.
\end{definition}

\begin{proposition}[Performance Error Bound]
\label{prop1}
\textit{If a partially symmetric Markov game $\mathcal{M}_g$  satisfies the conditions in equations \eqref{rewardequ} and \eqref{transequ} for all $(s, a) \in \Psi$, then the performance error $\textit{Error}_{\mathcal{M}_g}=|Q^{\star}(s, a) - Q^{\star}(gs, ga)|$ is bounded by $\frac{\epsilon}{1-\gamma} + \frac{\gamma \delta}{1-\gamma}$.}
\end{proposition} 
As stated in Prop~1, for the Partially Symmetric Markov game $\mathcal{M}_g$, the error introduced by incorporating symmetry samples is bounded. The proof of Prop~1 can be found in Section 1 of the Appendix\footnote{Video demonstrations and Supplementary materials are available at the project website
https://xinyu-site.github.io/PSE/.}. Prop~1 implies that the symmetry-augmented data are useful in MARL to a bounded extent even in partial symmetry situations like $\mathcal{M}_g$.

\section{Framework of the Partial Symmetry Exploitation}

%In this section, we present a general framework for exploiting symmetry prior, referred to as PSE.
This paper focuses on solving the following problem: \textit{In the context of partially symmetric Markov game $\mathcal{M}_g$, how can we appropriately leverage the symmetry prior to improving sample efficiency and performance of MARL?} On top of Prop~1, we propose a general framework, called Partial Symmetry Exploitation (PSE), for exploiting the symmetry prior, properly. The PSE framework is designed to adaptively utilize symmetry and is composed of four key modules: Symmetry Quantification, Adaptive Tuning, Symmetry Augmentation, and Symmetry Consistency Loss.

\subsection{Symmetry Quantification}
We propose a quantification method to measure the symmetry in $\mathcal{M}_g$. This method is applicable to various symmetries inherent in multi-agent systems, including permutation invariance, rotational equivariance, and translational invariance. We employ a transformed environment to assess the degree of symmetry by comparing the system's responses in both the original and transformed environments. For a partially symmetric Markov game $\mathcal{M}_g$, consider $(s, a, s')$ with an associated transformation $g$. In the transformed environment, action $ga$ is applied to the state $gs$, leading to a new state $\bar{s}^{\prime}$. We define a function $ D(gs', \bar{s}^{\prime}) $ to measure the degree of symmetry in MARL:
\begin{equation}
\begin{aligned}
D(gs', \bar{s}^{\prime}) & =1- \frac{1}{2} \frac{\|gs'-\bar{s}^{\prime}\|_2^2}{\|gs'\|_2^2+\| \bar{s}^{\prime}\|_2^2}, \\
\end{aligned}\label{distance}
\end{equation}
where the numerator represents the Euclidean norm of the difference between the vectors, and the denominator is the sum of individual Euclidean norms. Given a vector $ \mathbf{v} $, its Euclidean norm is defined as $\| \mathbf{v} \|_2 = \sqrt{\sum_i v_i^2}$. $D$ is a scalar and its values lie within the interval $[0, 1]$. A value of 1 indicates perfect symmetry and the lower bound of $D$ is attained when $gs'=-\bar{s}^{\prime}$. While we utilize the Euclidean norm in this context, other norms can also be employed.

Upon obtaining the measure of symmetry, we can introduce threshold values to determine the level of symmetry, categorizing the environment into Partial Symmetry (C1) and Non-Symmetry (C2). It's worth noting that perfect symmetry is a special case of partial symmetry. The threshold can be tuned according to the specific requirements of the problem and the performance trade-offs acceptable. 

\subsection{Adaptive Tuning}
In the early stages of training in $\mathcal{M}_g$, symmetry can assist the model to converge more swiftly and reduce the loss faster. However, as the model progressively adapts to the training environment and starts to capture more nuanced features of the data, an over-reliance on symmetry might have a negative effect on the model's training. To tackle this issue, as training advances, our PSE gradually reduces the dependence on symmetry. To this end, we present the following function:
\begin{equation}
\lambda (D,k) = D e^{-\beta k}, \label{lambdaa}
\end{equation}
which serves as the annealing coefficient at the $ k^{th}$ iteration, with $D$ signifying the degree of symmetry in Eq. (3) and $\beta$ denoting the decay rate. In the follow-up stage, as will be discussed later, the coefficient $ \lambda (D,k) $ serves as both  1) a probability to decide on whether the symmetry-augmented data is used and 2) a coefficient in the objective function to weigh the component of symmetry constraints. This auto-tuning approach strikes an adaptive balance on the extent to which the symmetry is leveraged in different training phases.

\subsection{Symmetry Augmentation}

One straightforward way to leverage symmetry is through data augmentation. Motivated by Prop 1, we present a data augmentation strategy designed to adaptively leverage symmetry. Based on Eq.~\eqref{lambdaa}, we obtain a coefficient $ \lambda_1 =D e^{-\beta_1 k}$ that starts with a value equal to the degree of symmetry and decreases over training iterations. Specifically, the coefficient $ \lambda_1 $ acts as a probabilistic threshold. If a random number $ r $ drawn from a uniform distribution between [0,1] is less than $ \lambda_1 $, data augmentation is applied in that iteration of training. This strategy ensures an expedited training process in the early phases. As training progresses, reliance on symmetry-augmented samples is reduced, thereby mitigating potential performance errors they might introduce.
\subsection{Symmetry Consistency Loss}

In multi-agent settings, using data augmentation to improve sample efficiency can be challenging. The reason is that when multiple agents are considered, more sources of variance are introduced, making the training unstable. The proposed symmetry consistency loss provides mitigation to this challenge (see Section 2 in the Appendix for more details). For a clean presentation, the MAPPO is used as an example to introduce symmetry consistency loss.

\textbf{Symmetry Consistency Loss}.
The policy consistency loss term $S_\pi(\theta)$ is defined as
\begin{equation}
S_\pi=K L\left[\pi_\theta(ga \mid gs) \mid \pi_\theta(a \mid s)\right],\label{symloss1}
\end{equation}
aiming to constrain distribution $\pi_\theta(ga \mid gs)$ to be close to $\pi_\theta(a \mid s)$. This helps guide the training process according to the symmetry prior. Assume that $V_\psi(s)$ represents an approximate value for state $s$, the symmetry consistency loss for value function is designed as
\begin{equation}
S_V=\mathbb{E}_{s}\left[\left(V_\psi(s)-V_\psi\left(gs\right)\right)^2\right],\label{symloss2}
\end{equation}
designed to minimize the discrepancy between the outputs of the value function when provided with the original input and the symmetry-transformed input. Therefore, we regard Eqs.~\eqref{symloss1} and \eqref{symloss2} as the symmetry consistency loss.

\textbf{MARL with Symmetry Consistency Loss}. Rather than having a fixed coefficient in the loss function, we utilize $\lambda _2=D e^{-\beta_2 k}$ calculated by Eq.~\eqref{lambdaa} to dynamically adjust the coefficient of symmetric loss. For the MAPPO, our PSE method optimizes the following loss objective:
\begin{equation}
J_{PSE} = J_{MAPPO} - \lambda_2 (S_\pi + S_V).
\end{equation}
The $J_{MAPPO}$ is the objective function for MAPPO and can be found in Section 2 of the Appendix.

\section{Experiments}

This section demonstrates our PSE's superiority via experiments in both simulated tasks and real-world robot systems.

\subsection{Environmental Settings}

\begin{figure}[ht]
     \centering
     \begin{subfigure}[b]{0.155\textwidth}        
         \includegraphics[width=\textwidth]{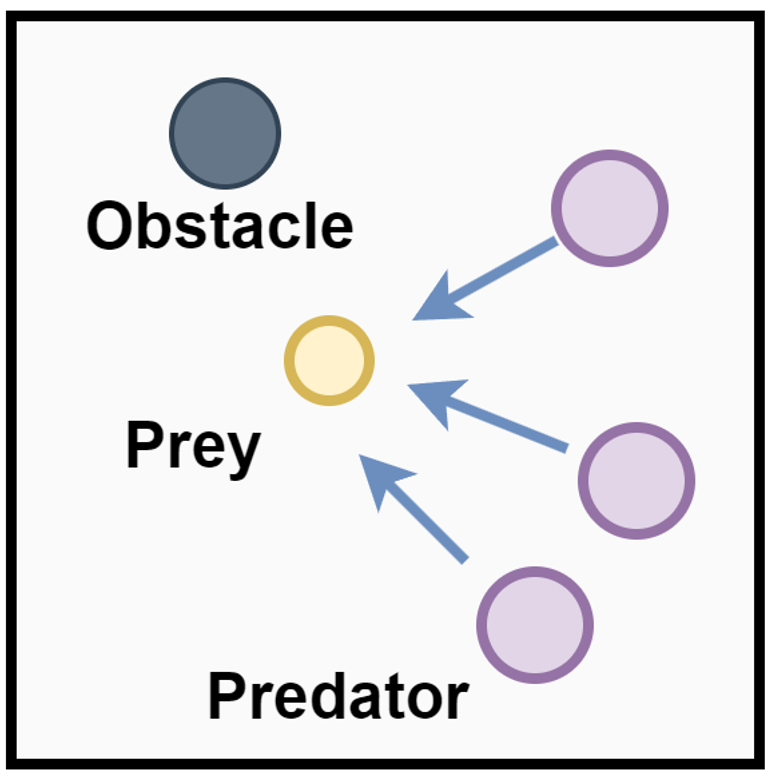}
         \caption{Predator-Prey}
         \label{schem1}
     \end{subfigure}    
     \begin{subfigure}[b]{0.155\textwidth}       
         \includegraphics[width=\textwidth]{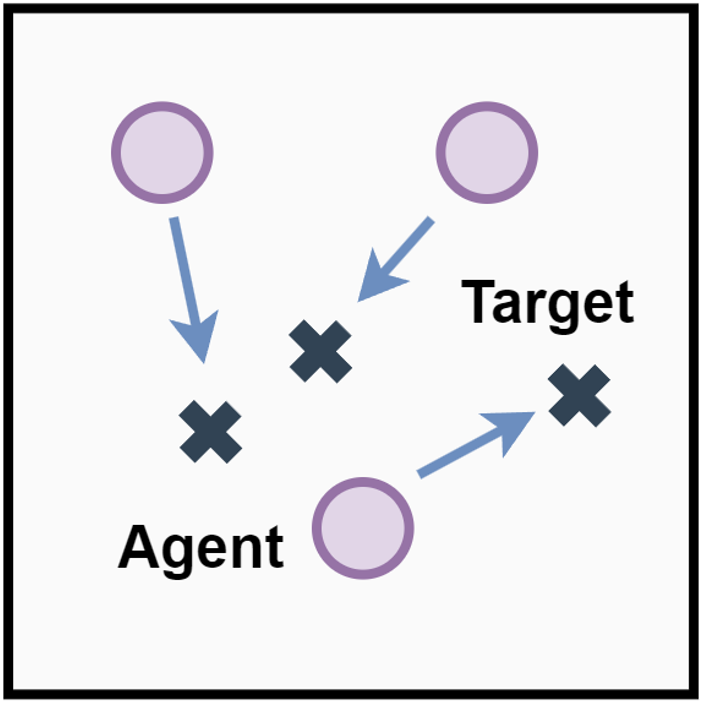}
         \caption{Navigation}
         \label{schem2}
     \end{subfigure}  
     \begin{subfigure}[b]{0.154\textwidth}          
         \includegraphics[width=\textwidth]{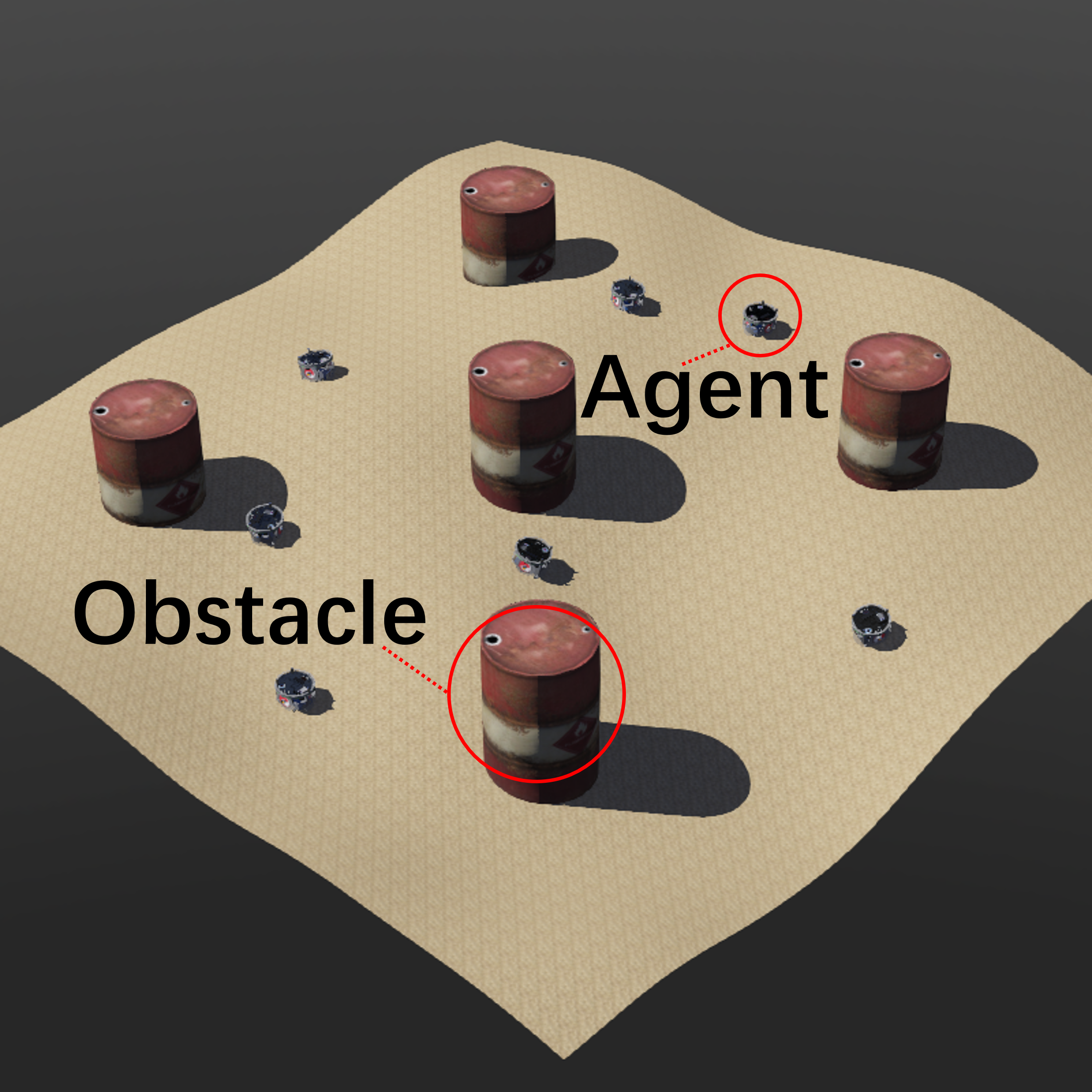}
         \caption{Formation change}
         \label{schem3}
     \end{subfigure}
     \caption{The simulated tasks considered in the experiments.}
     \label{schem}
\end{figure}
We conducted experiments in several tasks, including Predator-Prey (PP), Cooperative Navigation (CN), Wildlife Monitoring, and Formation Change (FC). CN and PP is a classic scenario implemented in multi-agent particle environment~\cite{MPE}. The wildlife monitoring is a grid-world-based environment, where a set of drones has to coordinate to accomplish the task~\cite{homon}. The goal is to trap poachers by having one drone hover above them while the other assists from the side. The details of FC are provided in Section 3 of the Appendix.  Figure \ref{schem} shows parts of the tasks.

\begin{figure*}[!t]
     \centering
     \subfloat[Predator-Prey]{%
         \includegraphics[width=0.32\textwidth]{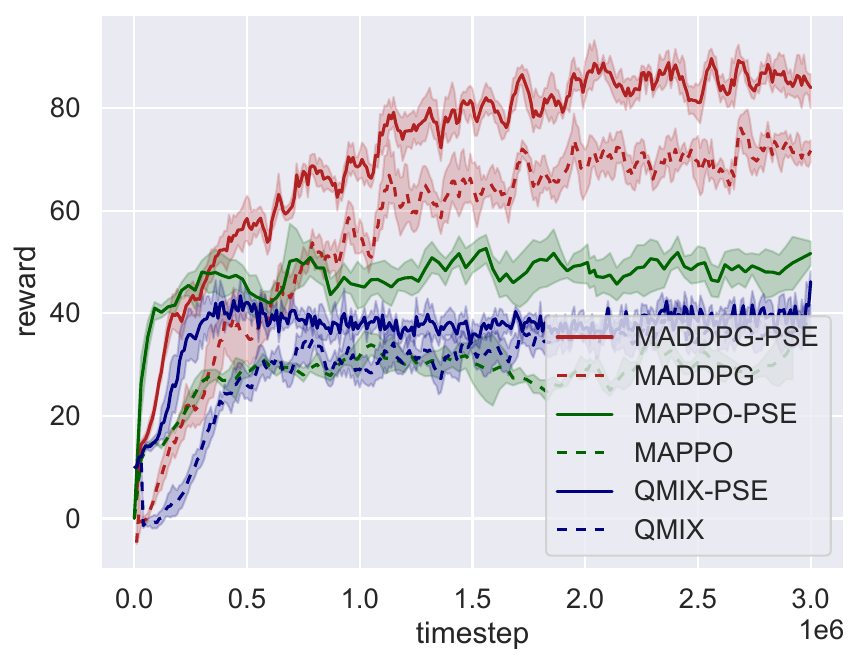}\label{5a}}
     \hfill
     \subfloat[Cooperative Navigation]{%
         \includegraphics[width=0.32\textwidth]{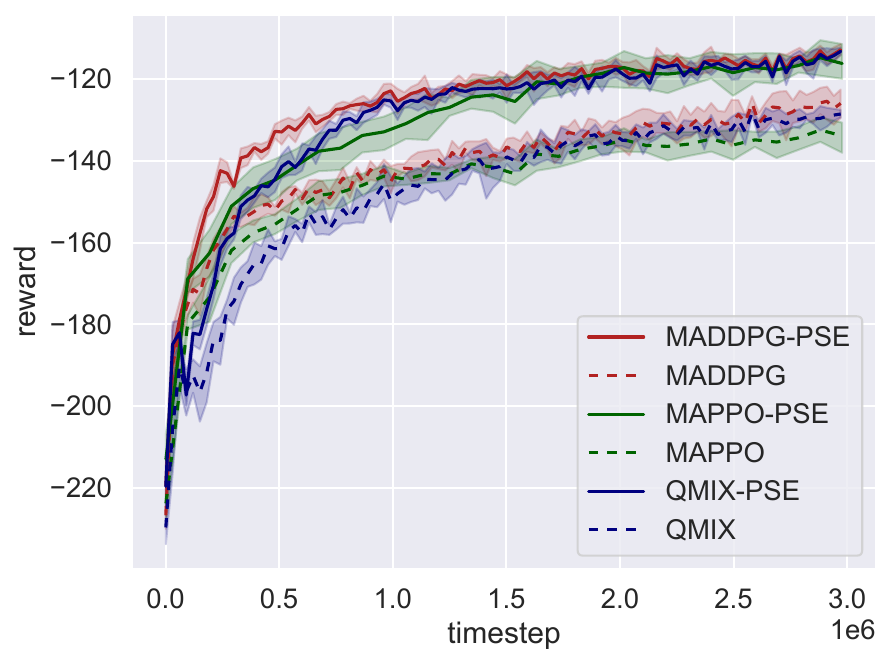}\label{5b}}
     \hfill
     \subfloat[Formation Change]{%
         \includegraphics[width=0.32\textwidth]{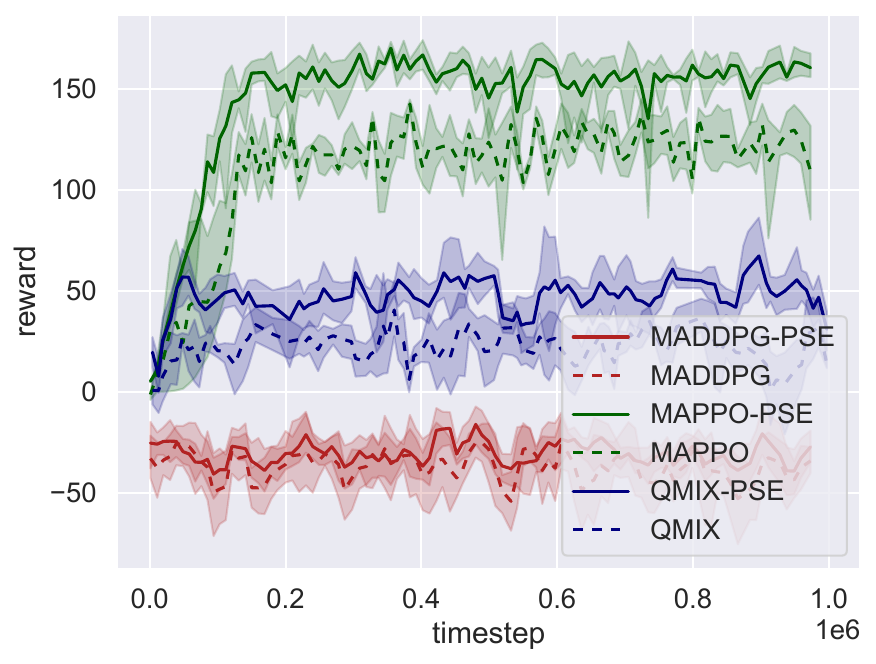}\label{5c}}
        \caption{Learning curves of the baseline and their versions with the PSE framework on the three multi-agent tasks.}
        \label{5}
\end{figure*}

\begin{figure*}[!t]
     \centering
     \begin{subfigure}[b]{0.32\textwidth}
         \centering
         \includegraphics[width=\textwidth]{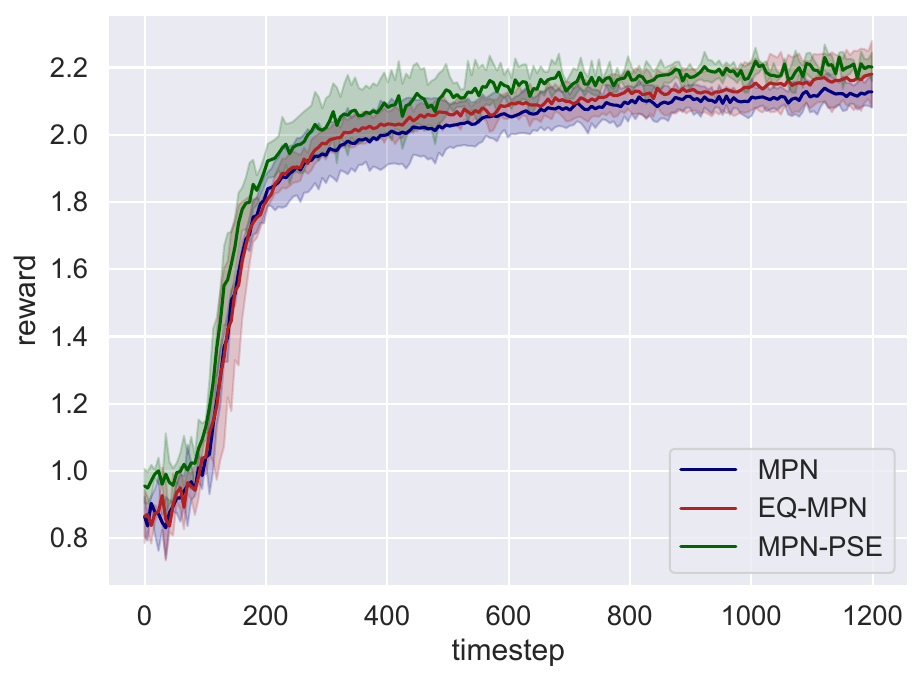}
         \caption{Noise Intensity Level 0}\label{6a}
     \end{subfigure}
     \hfill
     \begin{subfigure}[b]{0.32\textwidth}
         \centering
         \includegraphics[width=\textwidth]{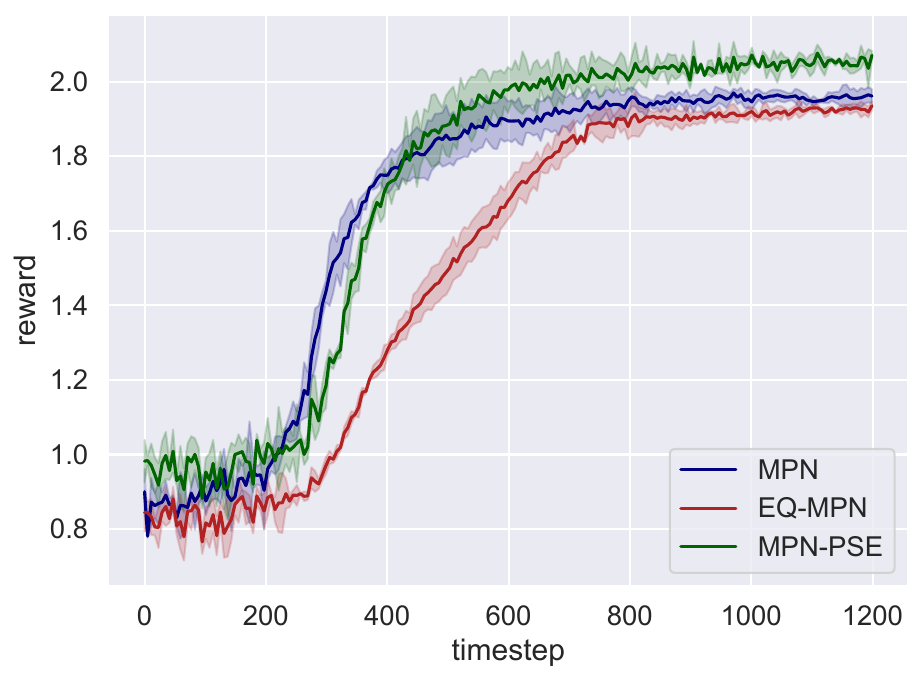}
         \caption{Noise Intensity Level 4}\label{6b}
     \end{subfigure}
     \hfill
     \begin{subfigure}[b]{0.32\textwidth}
         \centering
         \includegraphics[width=\textwidth]{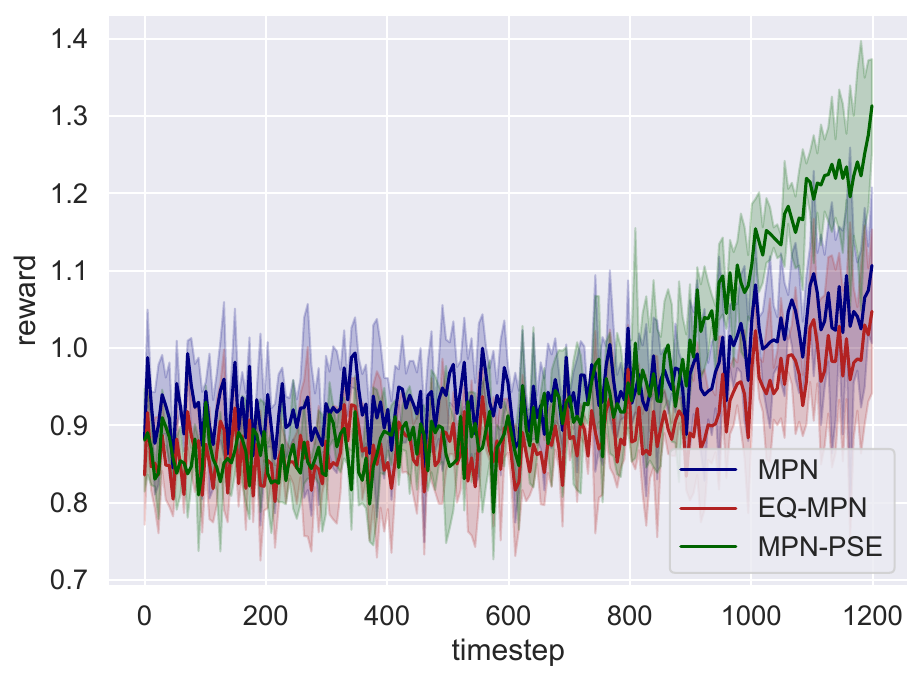}
         \caption{Noise Intensity Level 8}\label{6c}
     \end{subfigure}
     \caption{The learning curve of MPN, EQ-MPN and MPN-PSE under different symmetry-breaking in the Wildlife Monitoring.}\label{6}
\end{figure*}

\textbf{Symmetry Breaking.}  The original classic setups of these four tasks were designed to have perfect spatial symmetry. We deliberately modified these tasks to incorporate partial symmetry by integrating noise into the transition dynamics. The multi-robot FC task was conducted in the high-precision robot simulation environment Webots. As shown in Figure \ref{schem3}, robots were required to learn to avoid each other as well as the obstacles and coordinate to reach their destinations. We set an uneven terrain for this task to simulate partial symmetry. The severity of these uneven conditions can be modified, allowing us to explore the influences of environmental uncertainties on MARL in depth. For more details on symmetry breaking, please see Section 4 of the Appendix.

\textbf{Baselines}. The proposed PSE framework was applied to several baselines, including Multi-Agent Deep Deterministic Policy Gradient (MADDPG), Monotonic Value Function Factorisation for Deep MARL (QMIX), and Multi-Agent Proximal Policy Optimization (MAPPO), which are mainstream MARL approaches~\cite{qmix,maddpg,mappo}.

% iclr论文不能用在本文的三个任务上

\subsection{Main results}

This section presents the experimental results obtained using the setup described in Section 6.1. The performance of each algorithm was evaluated with 10 different random seeds, and the final experimental results under partial symmetry are shown in Figure \ref{5}. The results show that the MARL algorithms adopting the PSE framework achieved different degrees of advantage over their original versions.

\textbf{Predator-Prey.} In this scenario, there were three predators and one prey. As shown in Figure \ref{5a}, the proposed PSE framework outperformed the baseline methods significantly. The results indicated that the proposed framework could improve the data efficiency, convergence speed, and performance in terms of evaluation rewards.

\textbf{Cooperative Navigation.} The cooperative navigation was a fully cooperative environment, where 3 agents (circles) cooperated to reach 3 landmarks (crosses) under a minimum number of collisions. Similarly, as shown in Figure \ref{5b}, the results show that the proposed framework can improve data efficiency and performance in this task.

\begin{figure*}[!t]
    \centering
    \includegraphics[width=\textwidth]{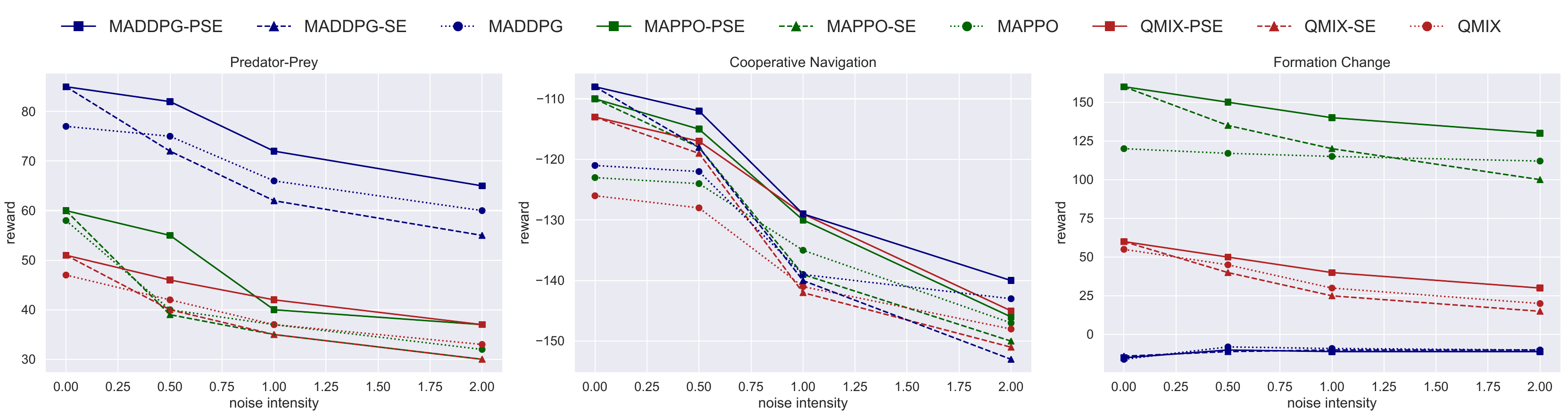}
    \caption{Convergence reward under varying noise intensities of various models across PP, CN, and FC scenarios.}\label{7}
\end{figure*}

\textbf{Formation Change.} To evaluate the proposed method in complex tasks, experiments were conducted on the multi-robot formation change task in the Webots simulator, as shown in Figure \ref{schem3}. In this scenario, 8 robots started in a square formation and had their destinations set on the opposite side. The experimental results show that the MADDPG and QMIX could not learn a useful policy in this task, whereas the agents trained by the MAPPO-PSE and MAPPO could reach the destination while avoiding collisions with each other and obstacles. As presented in Figure \ref{5c}, the algorithms enhanced by the proposed framework obtained higher rewards than the original versions. This indicates that the proposed PSE can further improve the performance of MARL algorithms in challenging environments.

\textbf{Wildlife Monitoring. } 
We conducted a comparison of three graph network-based methods: the Message Passing Network (MPN, a classic graph convolutional network), the EQ-MPN (an advanced baseline network embedding perfect symmetry as proposed in \cite{vand}) and MPN  with the PSE framework (MPN-PSE). The Wildlife Monitoring is specially analyzed here because it is the chosen environment in the literature well suited to the EQ-MPN, featuring pixel-based and grid-world states. Advanced results are achieved by EQ-MPN when the perfect symmetry in the environment holds. Here, we assessed the three models' performance across varying degrees of symmetry-breaking. Under conditions of perfect symmetry (see Figure 6a), the EQ-MPN with the perfect symmetry prior embedded in its network structure demonstrates superior performance compared to the classic MPN. However, our PSE framework surpasses both EQ-MPN and MPN in terms of convergence speed and the quality of the final convergence. As shown in Figure 6b, under partial symmetry, the performance of the EQ-MPN deteriorates, even to a level worse than the classic MPN. In contrast, our PSE-based method continues to enhance the performance of MPN. In Figure 6c, where the environment is completely devoid of any symmetry, all three methods face challenges in learning an acceptable policy. Yet, our PSE framework still maintains a discernible advantage.

\subsection{Impact of Different Degrees of Symmetry and Ablation Analysis }

We analyzed four distinct algorithm variations in our study: 1) MAPPO, which stands for the most primitive version of the algorithm. 2) MAPPO-SE, which retains only our symmetry augmentation and loss function components within the original MAPPO, and where the coefficient of the loss function is fixed to 0.5. 3) the MAPPO-PSE, our comprehensive framework representing the entirety of our proposed enhancements. Figure \ref{7} denotes each algorithm type with a distinct color, and different algorithm variations are highlighted by varying line types. It is observed that the PSE framework consistently excels across different degrees of symmetry-breaking. Interestingly, MAPPO-SE, which sticks to leveraging the perfect symmetry, experiences a substantial performance decline as noise intensity increases, even deteriorating to a level worse than the classic MAPPO. 

The results provide two insights: 1) a strong dependency on embedding perfect symmetry may seriously hamper the training and the final performance when the symmetry keeps breaking, and 2) our PSE framework can adapt to various symmetry-breaking conditions and consistently enhance the performance of mainstream multi-agent algorithms. The PSE enjoys this advantage due to the framework's symmetry quantification and adaptive tuning components. The same experiments are also conducted based on QMIX and MADDPG, which are included in Figure \ref{7}, and similar observations and conclusions can be obtained. The exception is that the MADDPG does not perform well in the FC task.

\subsection{Real world experiments}

\begin{figure}[!t]
     \centering
     \begin{subfigure}[b]{0.154\textwidth}
         \centering
         \includegraphics[width=\textwidth]{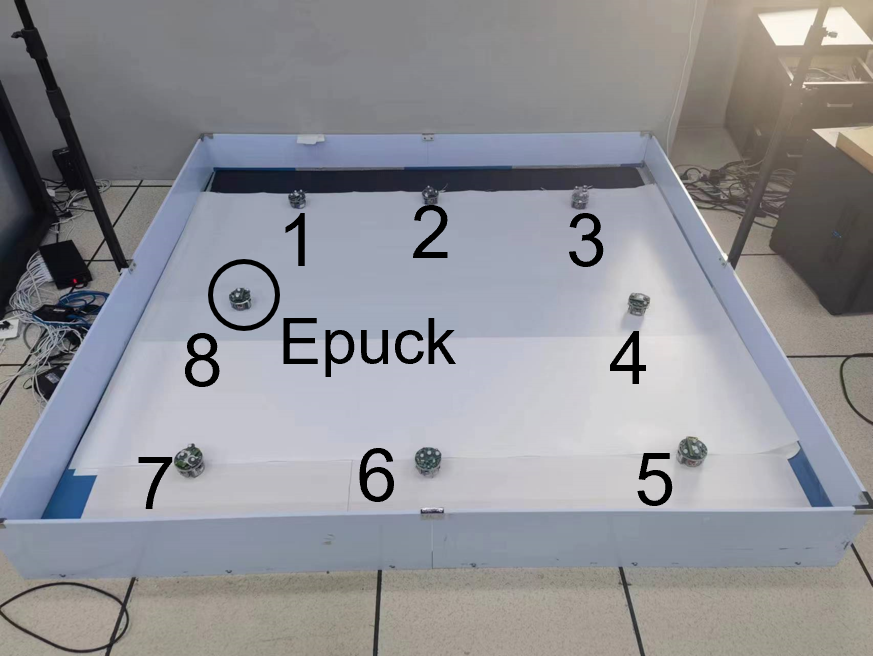}
         \caption{Start points}
         \label{fig:y equals x}
     \end{subfigure}
     \hfill
     \begin{subfigure}[b]{0.154\textwidth}
         \centering
         \includegraphics[width=\textwidth]{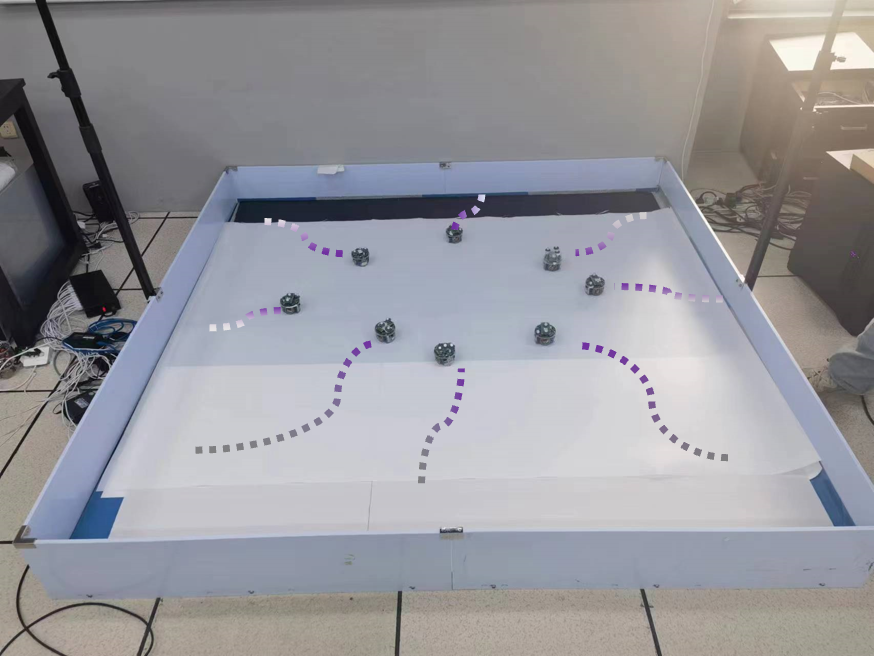}
         \caption{Trajectories}
         \label{fig:three sin x}
     \end{subfigure}
     \hfill
     \begin{subfigure}[b]{0.154\textwidth}
         \centering
         \includegraphics[width=\textwidth]{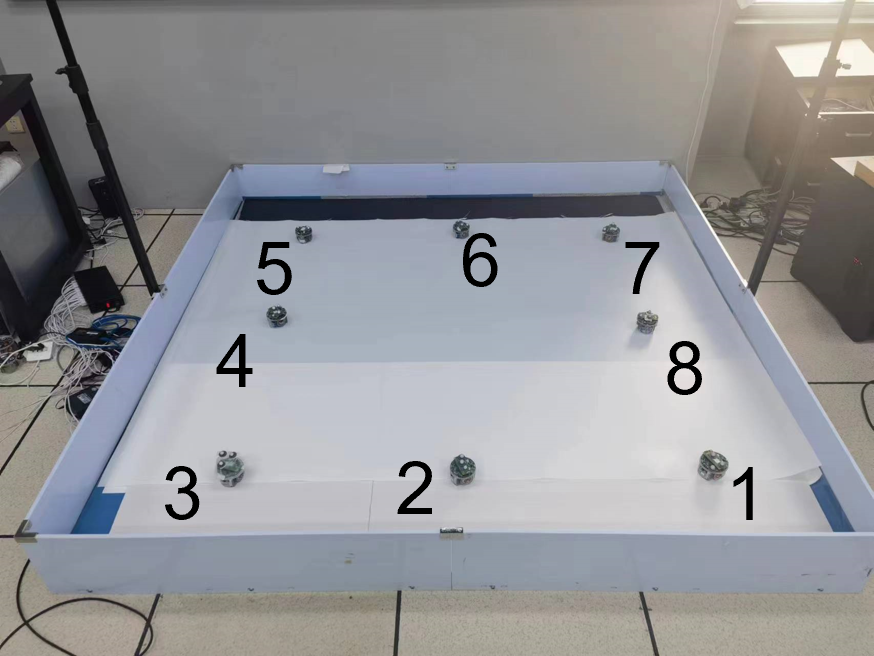}
         \caption{End points}
         \label{fig:five over x}
     \end{subfigure}
        \caption{Real-world formation change on a swarm of robots. The robots successfully switched their positions to the antipodal points by achieving collision avoidance.}
     \label{fig8}
\end{figure}

As shown in Figure \ref{fig8}, the real-world version of formation change presented in Section 6.1 was considered in this experiment. The trained policies were deployed on the Epuck, which is a small, lightweight robot platform. We followed a direct sim2real paradigm to deploy the policy network~\cite{s2rphy}. By incorporating our PSE approach into the MAPPO algorithm, the agents are able to complete tasks with fewer risky states. Risky states are defined as those in which the distance between agents is less than 5 centimeters, and the rate of risky states is the proportion of risky states to all states. The rate of risky states for MAPPO-PSE is 2.1\%, while the rate for MAPPO is 5.6\%. Details are provided in Section 5 of the Appendix.

\section{Conclusion}

In this paper, we newly introduce the partially symmetric Markov game. We then theoretically show that the corresponding performance error is bounded. Based on the bounded property, we propose a novel PSE framework to adaptively leverage symmetry prior in MARL. Experimental results support the superiority of PSE over baselines. In the future, we plan to extend the PSE to systems with heterogeneous agents, whose sensitivity to the symmetry-breaking conditions is different.

\section*{Acknowledgments}
This work was supported by the National Key Research and Development Program of China (Grant No. 2022ZD0117801), and the National Natural Science Foundation of China (Grant No. 62306023).

\bibliography{aaai24}

\begin{thebibliography}{27}
\providecommand{\natexlab}[1]{#1}

\bibitem[{Amadio, Colom{\'e}, and Torras(2019)}]{arm}
Amadio, F.; Colom{\'e}, A.; and Torras, C. 2019.
\newblock Exploiting symmetries in reinforcement learning of bimanual robotic
  tasks.
\newblock \emph{IEEE Robotics and Automation Letters}, 4(2): 1838--1845.

\bibitem[{Boutilier(1996)}]{Boutilier96}
Boutilier, C. 1996.
\newblock Planning, learning and coordination in multiagent decision processes.
\newblock In \emph{TARK}, volume~96, 195--210. Citeseer.

\bibitem[{Bronstein et~al.(2021)Bronstein, Bruna, Cohen, and
  Veli{\v{c}}kovi{\'c}}]{bros}
Bronstein, M.~M.; Bruna, J.; Cohen, T.; and Veli{\v{c}}kovi{\'c}, P. 2021.
\newblock Geometric deep learning: Grids, groups, graphs, geodesics, and
  gauges.
\newblock \emph{arXiv preprint arXiv:2104.13478}.

\bibitem[{De~Souza et~al.(2021)De~Souza, Newbury, Cosgun, Castillo, Vidolov,
  and Kuli{\'c}}]{s2rphy}
De~Souza, C.; Newbury, R.; Cosgun, A.; Castillo, P.; Vidolov, B.; and
  Kuli{\'c}, D. 2021.
\newblock Decentralized multi-agent pursuit using deep reinforcement learning.
\newblock \emph{IEEE Robotics and Automation Letters}, 6(3): 4552--4559.

\bibitem[{Feng et~al.(2023)Feng, Yu, Liang, Wu, and Tian}]{feng2023mact}
Feng, P.; Yu, X.; Liang, J.; Wu, W.; and Tian, Y. 2023.
\newblock MACT: Multi-agent Collision Avoidance with Continuous Transition
  Reinforcement Learning via Mixup.
\newblock In \emph{International Conference on Swarm Intelligence}, 74--85.
  Springer.

\bibitem[{Fillmore(1984)}]{rotationm}
Fillmore, J.~P. 1984.
\newblock A Note on Rotation Matrices.
\newblock \emph{IEEE Computer Graphics and Applications}, 4(2): 30--33.

\bibitem[{Gretton et~al.(2006)Gretton, Borgwardt, Rasch, Sch{\"o}lkopf, and
  Smola}]{MMD}
Gretton, A.; Borgwardt, K.; Rasch, M.; Sch{\"o}lkopf, B.; and Smola, A. 2006.
\newblock A kernel method for the two-sample-problem.
\newblock \emph{Advances in neural information processing systems}, 19.

\bibitem[{Jianye et~al.(2023)Jianye, Hao, Mao, Wang, Yang, Li, Zheng, and
  Wang}]{api}
Jianye, H.; Hao, X.; Mao, H.; Wang, W.; Yang, Y.; Li, D.; Zheng, Y.; and Wang,
  Z. 2023.
\newblock Boosting Multiagent Reinforcement Learning via Permutation Invariant
  and Permutation Equivariant Networks.
\newblock In \emph{The Eleventh International Conference on Learning
  Representations}.

\bibitem[{Laskin et~al.(2020)Laskin, Lee, Stooke, Pinto, Abbeel, and
  Srinivas}]{rlaug}
Laskin, M.; Lee, K.; Stooke, A.; Pinto, L.; Abbeel, P.; and Srinivas, A. 2020.
\newblock Reinforcement Learning with Augmented Data.
\newblock \emph{Advances in Neural Information Processing Systems}, 33.

\bibitem[{Laskin, Srinivas, and Abbeel(2020)}]{curl}
Laskin, M.; Srinivas, A.; and Abbeel, P. 2020.
\newblock Curl: Contrastive unsupervised representations for reinforcement
  learning.
\newblock In \emph{International Conference on Machine Learning}, 5639--5650.
  PMLR.

\bibitem[{Li et~al.(2023)Li, Guo, Xiu, Yu, Wang, Liu, Yang, and
  Liu}]{li2023byzantine}
Li, S.; Guo, J.; Xiu, J.; Yu, X.; Wang, J.; Liu, A.; Yang, Y.; and Liu, X.
  2023.
\newblock Byzantine Robust Cooperative Multi-Agent Reinforcement Learning as a
  Bayesian Game.
\newblock \emph{arXiv preprint arXiv:2305.12872}.

\bibitem[{Lin et~al.(2020)Lin, Huang, Zimmer, Guan, Rojas, and Weng}]{ITER}
Lin, Y.; Huang, J.; Zimmer, M.; Guan, Y.; Rojas, J.; and Weng, P. 2020.
\newblock Invariant transform experience replay: Data augmentation for deep
  reinforcement learning.
\newblock \emph{IEEE Robotics and Automation Letters}, 5(4): 6615--6622.

\bibitem[{Lowe et~al.(2017)Lowe, Wu, Tamar, Harb, Abbeel, and
  Mordatch}]{maddpg}
Lowe, R.; Wu, Y.; Tamar, A.; Harb, J.; Abbeel, P.; and Mordatch, I. 2017.
\newblock Multi-agent actor-critic for mixed cooperative-competitive
  environments.
\newblock \emph{arXiv preprint arXiv:1706.02275}.

\bibitem[{Mordatch and Abbeel(2017)}]{MPE}
Mordatch, I.; and Abbeel, P. 2017.
\newblock Emergence of Grounded Compositional Language in Multi-Agent
  Populations.
\newblock \emph{arXiv preprint arXiv:1703.04908}.

\bibitem[{Rashid et~al.(2018)Rashid, Samvelyan, Schroeder, Farquhar, Foerster,
  and Whiteson}]{qmix}
Rashid, T.; Samvelyan, M.; Schroeder, C.; Farquhar, G.; Foerster, J.; and
  Whiteson, S. 2018.
\newblock Qmix: Monotonic value function factorisation for deep multi-agent
  reinforcement learning.
\newblock In \emph{International Conference on Machine Learning}, 4295--4304.
  PMLR.
\newblock ISBN 2640-3498.

\bibitem[{Ravindran and Barto(2001)}]{report}
Ravindran, B.; and Barto, A.~G. 2001.
\newblock Symmetries and Model Minimization in Markov Decision Processes.
\newblock Technical report, University of Massachusetts, Amherst, MA, United
  States.

\bibitem[{Shi, Mo, and Di(2021)}]{shi2021physics}
Shi, R.; Mo, Z.; and Di, X. 2021.
\newblock Physics-informed deep learning for traffic state estimation: A hybrid
  paradigm informed by second-order traffic models.
\newblock In \emph{Proceedings of the AAAI Conference on Artificial
  Intelligence}, volume~35, 540--547.

\bibitem[{Shi et~al.(2022)Shi, Mo, Huang, Di, and Du}]{shi2021tits}
Shi, R.; Mo, Z.; Huang, K.; Di, X.; and Du, Q. 2022.
\newblock A physics-informed deep learning paradigm for traffic state and
  fundamental diagram estimation.
\newblock \emph{IEEE Transactions on Intelligent Transportation Systems}, 23:
  11688--11698.

\bibitem[{Shi, Steenkiste, and Veloso(2021)}]{shi2021improving}
Shi, R.; Steenkiste, P.; and Veloso, M.~M. 2021.
\newblock Improving the on-vehicle experience of passengers through SC-M*: A
  scalable multi-passenger multi-criteria mobility planner.
\newblock \emph{IEEE Transactions on Intelligent Transportation Systems},
  22(2): 1026--1040.

\bibitem[{van~der Pol et~al.(2021)van~der Pol, van Hoof, Oliehoek, and
  Welling}]{vand}
van~der Pol, E.; van Hoof, H.; Oliehoek, F.~A.; and Welling, M. 2021.
\newblock Multi-Agent MDP Homomorphic Networks.
\newblock \emph{arXiv preprint arXiv:2110.04495}.

\bibitem[{van~der Pol et~al.(2020)van~der Pol, Worrall, van Hoof, Oliehoek, and
  Welling}]{homon}
van~der Pol, E.; Worrall, D.; van Hoof, H.; Oliehoek, F.; and Welling, M. 2020.
\newblock MDP homomorphic networks: Group symmetries in reinforcement learning.
\newblock \emph{Advances in Neural Information Processing Systems}, 33.

\bibitem[{Wang, Walters, and Platt(2022)}]{so2}
Wang, D.; Walters, R.; and Platt, R. 2022.
\newblock $\mathrm{SO}(2)$-Equivariant Reinforcement Learning.
\newblock \emph{arXiv preprint arXiv:2203.04439}.

\bibitem[{Yarats, Kostrikov, and Fergus(2020)}]{imageallneed}
Yarats, D.; Kostrikov, I.; and Fergus, R. 2020.
\newblock Image augmentation is all you need: Regularizing deep reinforcement
  learning from pixels.
\newblock In \emph{International Conference on Learning Representations}.

\bibitem[{Ye et~al.(2021)Ye, Chen, Jiang, Song, Yang, and Fan}]{homoaug}
Ye, Z.; Chen, Y.; Jiang, X.; Song, G.; Yang, B.; and Fan, S. 2021.
\newblock Improving sample efficiency in Multi-Agent Actor-Critic methods.
\newblock \emph{Applied Intelligence}, 1--14.

\bibitem[{Yu et~al.(2021{\natexlab{a}})Yu, Velu, Vinitsky, Wang, Bayen, and
  Wu}]{mappo}
Yu, C.; Velu, A.; Vinitsky, E.; Wang, Y.; Bayen, A.; and Wu, Y.
  2021{\natexlab{a}}.
\newblock The Surprising Effectiveness of PPO in Cooperative, Multi-Agent
  Games.
\newblock \emph{arXiv preprint arXiv:2103.01955}.

\bibitem[{Yu et~al.(2023)Yu, Shi, Feng, Tian, Luo, and Wu}]{yu2023esp}
Yu, X.; Shi, R.; Feng, P.; Tian, Y.; Luo, J.; and Wu, W. 2023.
\newblock ESP: Exploiting Symmetry Prior for Multi-Agent Reinforcement
  Learning.
\newblock In \emph{ECAI 2023}, 2946--2953. IOS Press.

\bibitem[{Yu et~al.(2021{\natexlab{b}})Yu, Wu, Feng, and Tian}]{yu2021swarm}
Yu, X.; Wu, W.; Feng, P.; and Tian, Y. 2021{\natexlab{b}}.
\newblock Swarm inverse reinforcement learning for biological systems.
\newblock In \emph{2021 IEEE International Conference on Bioinformatics and
  Biomedicine (BIBM)}, 274--279. IEEE.

\end{thebibliography}

\ifarXiv
    \foreach \x in {1,...,\numbersupplementpages}
    {
        \ifodd\x
            \includepdf[pages={\x},noautoscale=true,pagecommand={},offset=0cm 0cm]{\supplementfilename}
        \else
            \includepdf[pages={\x},noautoscale=true,pagecommand={},offset=0cm 0cm]{\supplementfilename}
        \fi
    }
\fi

\end{document}